\documentclass[prd,byrevtex
,preprint
,showkeys
,showpacs]{revtex4}
\usepackage{amsmath,amsfonts}
\usepackage[dvipdfm]{graphicx}
\usepackage{float}
\newcommand{\bs}[1]{\boldsymbol{#1}}
\newcommand{\pa}{\partial}
\newcommand{\al}{\alpha}
\newcommand{\del}{\delta}
\newcommand{\ep}{\epsilon}
\begin{document}
\title{Entanglement of Quantum Fluctuations in the Inflationary Universe}
\author{Yasusada Nambu}
\affiliation{Department of Physics, Graduate School of Science, Nagoya 
University, Chikusa, Nagoya 464-8602, Japan}
\email{nambu@gravity.phys.nagoya-u.ac.jp}

\date{July 15, 2008 ver 1.0}  
\begin{abstract}
  We investigate quantum entanglement of a scalar field in the
  inflationary universe. By introducing a bipartite system using a
  lattice model of scalar field, we apply the separability criterion
  based on the partial transpose operation and numerically calculate
  the bipartite entanglement between separate spatial regions. We find
  that the initial entangled state becomes separable or disentangled
  when the size of the spatial regions exceed the Hubble horizon. This
  is a necessary condition for the appearance of classicality of the
  quantum fluctuation. We further investigate the condition for the
  appearance of the classical distribution function and find that the
  condition is given by the inequality for the symplectic eigenvalue
  of the covariance matrix of the scalar field.
\end{abstract}
\maketitle
\section{Introduction}

Inflation provides a mechanism of generation of primordial
inhomogeneity which is needed for the formation of large scale
structures in our present universe.  During the qusi-de Sitter
expansion stage of the inflationary universe, short wavelength quantum
fluctuations of the inflaton field are generated by particle creations
and then they are stretched by the cosmic expansion and their
wavelength exceeds the Hubble horizon beyond which the physical
process proceeds independently.  After the wavelength of generated
quantum fluctuations become larger than the Hubble horizon, the
quantum nature of the fluctuation is expected to be lost and the
statistical property of fluctuations is replaced by the classical
distribution function. This is the assumption of the quantum to
classical transition of the quantum fluctuation generated during the
inflation. Once this assumption is adopted, we can use generated
``classical'' fluctuations as initial perturbations for the
large-scale structure formation, which obeys deterministic classical
dynamics.

We must explain or justify this assumption of the quantum to classical
transition of primordial fluctuation and many investigations have been
done on this subject so
far\cite{GuthAH:PRD32:1985,SakagamiM:PTP79:1988,BrandenbergerRH:MPLA5:1990,NambuY:PL276B:1992,AlbrechtA:PRD50:1994,PolarskiD:CQG13:1996,LesgourguesJ:NPB497:1997,KieferC:CQG15:1998,KieferC:CQG24:2007,SharmanJW:JCAP11:2007,MartineauP:CQG24:2007,BurgessCP:PRD77:2008}.
A rough outline of the classicalization mechanism proposed in
researches\cite{GuthAH:PRD32:1985,AlbrechtA:PRD50:1994,PolarskiD:CQG13:1996,LesgourguesJ:NPB497:1997,KieferC:CQG15:1998,KieferC:CQG24:2007}
is as follows: In the inflationary universe, particle creations occur
due to the accelerated expansion of the Universe and the quantum field
becomes a highly squeezed state.  For such a highly excited state as a
squeezed state, noncommutativity between canonically conjugate
variables can be neglected and the operator nature of variables are
effectively lost when we evaluate the expectation value of quantum
variables. This means we can regard operators as c-numbers.
Furthermore, it can be shown that there appears a sharp peak around a
line of the Wigner function for the state of quantum fluctuations and
this indicates establishment of classical correlation between
canonically conjugate variables on the phase space. At this stage, the
Wigner function itself can be interpreted as a classical probability
distribution function and the quantum fluctuations themselves can be
treated as classical stochastic variables.  Hence, after sufficient
squeezing, we can represent the nature of the quantum fluctuations by
the classical stochastic variables with an appropriate probability
distribution function of which property is determined by the state of
the quantum fluctuations. For the classicality of the quantum system,
we also need the mechanism of decoherence and the studies along this
line have been done with the assumption of an appropriate coupling
between the system and the
environment\cite{SakagamiM:PTP79:1988,BrandenbergerRH:MPLA5:1990,NambuY:PL276B:1992,SharmanJW:JCAP11:2007,MartineauP:CQG24:2007,BurgessCP:PRD77:2008}.

However, these analysis do not pay attention to an important aspect of
quantum mechanics, quantum entanglement, which definitely
distinguishes a quantum world from a classical one. When we calculate
a correlation function of observables between spatially separated two
regions, we have a possibility that the quantum correlation function
cannot be reproduced using a local classical probability distribution
function if these two regions are
entangled\cite{EinsteinA:PR47:1935,BellJS:P1:1964} and the classical
locality is violated. In other words, we cannot regard the quantum
fluctuations as the classical stochastic fluctuations as long as the
system is entangled. Therefore, it is important to clarify the
relation between the entanglement and the appearance of the classical
nature to fully understand the mechanism of the quantum to classical
transition of primordial fluctuations.

In this paper, we consider the entanglement of quantum field between
two spatially separated regions in a de Sitter universe and aim to
understand the mechanism of the quantum to classical transition of
fluctuation from the viewpoint of quantum entanglement.  This paper is
organized as follows: We introduce the concept of entanglement and
separability in Sec.~II. In Sec.~III, we calculate the bipartite
entanglement of a scalar field in the de Sitter universe. In Sec.~IV,
we discuss the relation between the entanglement and the quantum to classical
transition in the inflationary universe. Sec.~V is devoted to summary
and conclusion.  We use units in which $c=\hbar=8\pi G=1$ throughout
the paper.
\section{Separability and Entanglement}
In this paper, we focus on a bipartite system composed of
two Gaussian modes described by canonical variables $(\hat q_1,\hat
p_1)$ and $(\hat q_2,\hat p_2)$. This is the simplest case deriving
entanglement for continuous variable system. Let Alice be in
possession of mode 1 and let Bob be in possession of mode 2. A quantum
state $\hat \rho$ of the bipartite system is defined to be separable
if and only if $\hat\rho$ can be expressed in the
form\cite{BraunsteinSL:RMP77:2005}
\begin{equation}
  \hat\rho=\sum_jp_j\hat\rho_{jA}\otimes\hat\rho_{jB},\quad
  \sum_jp_j=1,\quad p_j\ge 0,
\end{equation}
where $\hat\rho_{jA}$ and $\hat\rho_{jB}$ are density operators of the
modes of Alice and Bob, respectively. If the state of the system
cannot be expressed in this form, the quantum state of the system is
called entangled. When the state is entangled, the observables
associated to the party A and B are correlated and their correlations
cannot be reproduced with purely classical means. This leads to the
phenomena peculiar to the quantum mechanics such as the EPR
correlation\cite{EinsteinA:PR47:1935} and the  violation of Bell's
inequality\cite{BellJS:P1:1964}.

For a bipartite Gaussian two mode system, we have necessary and
sufficient conditions for
separability\cite{SimonR:PRL84:2000,DuanL:PRL84:2000} and we can judge
 whether the system is entangled or not using these criteria. In this paper,
we adopt a criterion proposed by Simon\cite{SimonR:PRL84:2000} which
uses the partial transpose operation for a bipartite system. We define
the phase space variables as
\begin{equation}
  \hat{\bs{\xi}}=
  \begin{pmatrix}
    \hat q_1 \\ \hat p_1 \\ \hat q_2 \\ \hat p_2
  \end{pmatrix}
\end{equation}
Using these variables, the canonical commutation relations are expressed as
\begin{align}
 &\left[\hat\xi_\al,\hat\xi_\beta\right]=i\Omega_{\al\beta},\quad\al,\beta=1,2,3,4,
 \notag \\
 &\bs{\Omega}=
 \begin{pmatrix}
   \bs{J} & 0 \\ 0 & \bs{J}
 \end{pmatrix}
,\quad \bs{J}=
\begin{pmatrix}
  0 & 1 \\ -1 & 0
\end{pmatrix}.
\end{align}
The Gaussian state is completely characterized by the covariance matrix
\begin{equation}
  V_{\al\beta}=\frac{1}{2}
  \left\langle\hat\xi_\al\hat\xi_\beta+\hat\xi_\beta\hat\xi_\al\right\rangle
  =\frac{1}{2}\mathrm{tr}\left((\hat\xi_\al\hat\xi_\beta+\hat\xi_\beta\hat\xi_\al)
    \hat\rho\right) 
\end{equation}
where we assume the state with $\langle\xi_\al\rangle=0$. For a physical
state, the density matrix must be non-negative and the corresponding
covariance matrix must satisfy the inequality\cite{SimonR:PRL84:2000}
\begin{equation}
  \label{eq:positivity}
  \bs{V}+\frac{i}{2}\bs{\Omega}\ge 0
\end{equation}
which is the generalization of the uncertainty relation between two
canonically conjugate variables. The separability of the bipartite Gaussian state
is expressed in terms of the partial transpose operation defined by
\begin{equation}
  \hat{\bs{\xi}}'=\bs{\Lambda\hat\xi},\quad \bs{\Lambda}=\mathrm{diag}(1,1,1,-1).
\end{equation}
This operation reverses the sign of  Bob's momentum. With this
operation, the covariance matrix transforms as
\begin{equation}
  \tilde{\bs{V}}=\bs{\Lambda V\Lambda}^T.
\end{equation}
The necessary and sufficient condition of the separability is given by
the inequality
\begin{equation}
  \label{eq:separability}
  \tilde{\bs{V}}+\frac{i}{2}\bs{\Omega}\ge 0
\end{equation}
which represents the physical condition for the partially transposed
state. For an  entangled state, this inequality is violated and the
partially transposed state becomes unphysical.  To measure the degrees
of entanglement, we introduce the logarithmic negativity via
symplectic eigenvalues of the covariance matrix.  The covariance
matrix can be diagonalized by an appropriate symplectic transformation
$\bs{S}\in\mathrm{Sp}(4,R), \bs{S}\bs{\Omega}\bs{S}^T=\bs{\Omega}$ as
follows\cite{AdessoG:PRL92:2004,AdessoG:PRA70:2004}
\begin{equation}
  \bs{SVS}^T=\mathrm{diag}(\nu_{+},\nu_{+},\nu_{-},\nu_{-}),\quad
  \nu_{+}\ge\nu_{-}\ge 0
\end{equation}
where $\nu_{\pm}$ are symplectic eigenvalues. In terms of symplectic
eigenvalues, the physical condition \eqref{eq:positivity} can be
expressed as
\begin{equation}
  \label{eq:posi-symp}
  \nu_{-}\ge\frac{1}{2}
\end{equation}
and the separability condition \eqref{eq:separability} can be
expressed as
\begin{equation}
  \label{eq:sepa-symp}
  \tilde\nu_{-}\ge\frac{1}{2}.
\end{equation}
The logarithmic negativity is defined by
\begin{equation}
  \label{eq:lognegativity}
  E_N=-\mathrm{min}\left[\log_2(2\tilde\nu_{-}),0\right].
\end{equation}
For an entangled state, $\tilde\nu_{-}<1/2$ and we have $E_N>0$. For a
separable state, $\tilde\nu_{-}\ge 1/2$ and we have
$E_N=0$. Practically, the symplectic eigenvalues can be obtained as
eigenvalues of the matrix $i\bs{\Omega V}$\cite{AdessoG:PRL92:2004,AdessoG:PRA70:2004}.
 
\section{Entanglement of Quantum Field in the de Sitter universe}
\subsection{One-dimensional lattice model of scalar field}
To comprehend the behavior of the entanglement of quantum fields in
the inflationary universe, we consider a real massless scalar field
$\phi$ in the de Sitter universe. The metric and the Lagrangian are
\begin{align}
  &ds^2=a(\eta)^2(-d\eta^2+d\bm{x}^2),\quad a=-\frac{1}{H\eta},\quad -\infty<\eta<0,
  \notag\\
  &L=\int d^3x\sqrt{-g}\left(-\frac{1}{2}g^{\mu\nu}\pa_\mu\phi\pa_\nu\phi\right),
\end{align}
where $\eta$ is the conformal time and $H$ is the Hubble parameter, 
which is assumed to be constant in time.  By introducing a 
conformally rescaled variable $q=a\phi$,
\begin{equation}
  L=\int d^3x\frac{1}{2}\left[\left(q'-\frac{a'}{a}q\right)^2-(\pa_i q)^2\right],
\end{equation}
and the equations of motion of the scalar field is
\begin{equation}
  q''-\frac{a''}{a}q-\pa_i^2q=0,
\end{equation}
where $'$ denotes the derivative with respect to the conformal time
$\eta$. 

To investigate the property of the quantum entanglement of the scalar
field, we adopt a discrete lattice model of the scalar field in this
paper.  This model introduces a cutoff of short wavelength mode of the
scalar field, which regularizes the ultraviolet divergence of the
vacuum fluctuations. The same model is used to investigate the spatial
structure of entanglement in a Minkowski
spacetime\cite{KoflerJ:PRA73:2006}. To simplify the analysis, we
assume that the scalar field depends only on one spatial coordinate
and the space is one-dimensional. Then the lattice version of the
scalar field Lagrangian is
\begin{equation}
  L=\frac{\Delta
    x}{2}\sum_{j=1}^N\left[\left(q_j'-\frac{a'}{a}q_j\right)^2(\Delta
    x)^2-(q_j-q_{j-1})^2\right]
\end{equation}
where $q_j$ denotes the scalar field at the $j$-th lattice site, $\Delta
x$ is a lattice spacing and $N$ is the total number of lattice
sites. The equation of motion is
\begin{align}
  &q_j''-\frac{a''}{a}q_j+\frac{1}{(\Delta
    x)^2}\left[2q_j-\al(q_{j+1}+q_{j-1})\right]=0,
  \quad
  j=1,2,\cdots,N,\\
  & q_0=q_{N}, q_{N+1}=q_1 \notag
\end{align}
where we assume a periodic boundary condition for $q_j$, and the
parameter $\al\neq 1$ is introduced to regularize the infrared
divergence which appears in the correlation function of the scalar
field. This divergence is peculiar to one-dimensional massless scalar
field. The nonunity value of the parameter $\alpha$ corresponds to adding
a small mass to the scalar field
\begin{equation}
  m^2=\frac{2(1-\al)}{(\Delta x)^2}
\end{equation}
and we choose the value of $\alpha$ sufficiently close to unity so
that the our result of calculation does not depends on the value of
this cutoff parameter. By rescaling the time variable as
$\eta\rightarrow\eta \Delta x$, the equation of motion can be written
as
\begin{equation}
  q_j''-\frac{a''}{a}q_j+2q_j-\al(q_{j+1}+q_{j-1})=0.
\end{equation}
The Hamiltonian is
\begin{equation}
  H=\sum_{j=1}^N\left[\frac{1}{2}p_j^2+q_j^2-\al q_jq_{j-1}+\frac{a'}{a}p_jq_j\right].
\end{equation}
To quantize this system, we introduce the Fourier expansion of the
scalar field on the lattice as follows
\begin{equation}
  q_j=\frac{1}{\sqrt{N}}\sum_{k=0}^{N-1}\tilde q_k\,e^{i\theta_k
    j},\quad p_j=\frac{1}{\sqrt{N}}\sum_{k=0}^{N-1}\tilde p_k^*
  e^{i\theta_k j},\quad\theta_k=\frac{2\pi k}{N}.
\end{equation}
The equation of motion for the Fourier mode $\tilde q_k$ is
\begin{equation}
  \tilde q_k''+\left(\omega_k^2-\frac{a''}{a}\right)q_k=0,\quad 
  \omega_k^2=2(1-\al\cos\theta_k).
\end{equation}
Introducing creation and annihilation operators, the quantized canonical
variables are represented as follows
\begin{align}
  &\hat q_j=\frac{1}{\sqrt{N}}\sum_{k=0}^{N-1}\left(f_k\hat
    a_k+f_k^*\hat a_{N-k}^\dag\right)e^{i\theta_k j}, \\
  &\hat p_j=\frac{1}{\sqrt{N}}\sum_{k=0}^{N-1}(-i)\left(g_k\hat
    a_k-g_k^*\hat a_{N-k}^\dag\right)e^{i\theta_k j}, \\
 &[\hat q_j,\hat p_{\ell}]=i\delta_{j\ell},\quad
   [\hat a_{k_1}, \hat a_{k_2}{}^{\!\!\!\dag}]=\del_{k_1,k_2},\quad 
   [\hat a_{k_1}, \hat a_{k_2}]=[\hat a_{k_1}{}^{\!\!\!\dag}, \hat
   a_{k_2}{}^{\!\!\!\dag}]=0, \\
 & f_k''+\left(\omega_k^2-\frac{a''}{a}\right)f_k=0,\quad
 g_k=i\left(f_k'-\frac{a'}{a}f_k\right),\quad f_kf_k^*{}'-f_k'f_k^*=i.
\end{align}
As the quantum state of the scalar field, we assume the Bunch-Davis
vacuum state and it corresponds to the following form of the mode functions
\begin{equation}
  f_k=\frac{1}{\sqrt{2\omega_k}}\left(1+\frac{1}{i\omega_k\eta}\right)e^{-i\omega_k\eta},
\quad
  g_k=\sqrt{\frac{\omega_k}{2}}\,e^{-i\omega_k\eta}.
\end{equation}
The two point correlation functions between the canonical variables on the lattice
sites are given by
\begin{align}
  &g_{|j-\ell|}\equiv\frac{1}{2}\langle\hat q_j\hat q_\ell+\hat
  q_\ell\hat q_j\rangle=\frac{1}{N}\sum_{k=0}^{N-1}|f_k|^2\cos(\theta_k(j-\ell)),\\
  &h_{|j-\ell|}\equiv\frac{1}{2}\langle\hat p_j\hat p_\ell+\hat
  p_\ell\hat p_j\rangle=\frac{1}{N}\sum_{k=0}^{N-1}|g_k|^2\cos(\theta_k(i-\ell)),\\
  &k_{|j-\ell|}\equiv\frac{1}{2}\langle\hat q_j\hat p_\ell+\hat
  p_\ell\hat
  q_j\rangle=\frac{1}{N}\sum_{k=0}^{N-1}\frac{i}{2}(f_kg_k^*-f_k^*g_k)
\cos(\theta_k(j-\ell)).
\end{align}

\begin{figure}[H]
  \centering
  \includegraphics[width=0.6\linewidth,clip]{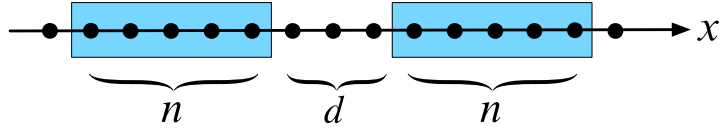}
  \caption{A bipartite system in our one-dimensional lattice model.}
  \label{fig:bipartite}
\end{figure}

Now, we define a bipartite system using this lattice model. As we are
interested in the correlation and the entanglement between different
spatial regions, we introduce the following block variables by
spatially averaging the variables in given regions A and B (see
Fig.~\ref{fig:bipartite}).
\begin{equation}
  \hat q_A=\frac{1}{\sqrt{n}}\sum_{j\in A}\hat q_j,\quad
  \hat p_A=\frac{1}{\sqrt{n}}\sum_{j\in A}\hat p_j,\quad
  \hat q_B=\frac{1}{\sqrt{n}}\sum_{j\in B}\hat q_j,\quad
  \hat p_B=\frac{1}{\sqrt{n}}\sum_{j\in B}\hat p_j.
\end{equation}
\noindent
The region A and B contain $n$ lattice sites and the coarse-grained
field values are assigned to each regions. The separation between A
and B is $d$.  The commutation relations between these coarse-grained
variables are
\begin{equation}
   [\hat q_A,\hat p_A]=[\hat q_B,\hat p_B]=i,\quad [\hat q_A,\hat
 p_B]=0
\end{equation}
and the set of canonical variables $(\hat q_A,\hat p_A,\hat q_B,\hat
p_B)$ constitutes a bipartite system.   The covariance matrix of this
bipartite system is given by the following symmetric $4\times 4$
matrix:
\begin{align}
  &\bs{V}=
  \begin{pmatrix}
    \bs{A} & \bs{C} \\ \bs{C} & \bs{A}
  \end{pmatrix}
,\quad \bs{A}=
\begin{pmatrix}
  a_1 & a_3 \\ a_3 & a_2
\end{pmatrix}
,\quad \bs{C}=
\begin{pmatrix}
  c_1 & c_3 \\ c_3 & c_2
\end{pmatrix}
, \\
 & a_1=\langle\hat q_A^2\rangle=\langle\hat q_B^2\rangle=\frac{1}{n}\sum_{i,j\in A}g_{|i-j|},
 \notag \\
 & a_2=\langle\hat p_A^2\rangle=\langle\hat p_B^2\rangle=\frac{1}{n}\sum_{i,j\in A}h_{|i-j|},
 \notag \\
 & a_3=\frac{1}{2}\langle\hat q_A\hat p_A+\hat p_A\hat q_A\rangle
=\frac{1}{n}\sum_{i,j\in A}k_{|i-j|}, \notag \\
 & c_1=\frac{1}{2}\langle\hat q_A\hat q_B+\hat q_B\hat q_A\rangle
   =\frac{1}{n}\sum_{i\in A, j\in B}g_{|i-j|}, \notag \\
& c_2=\frac{1}{2}\langle\hat p_A\hat p_B+\hat p_B\hat p_A\rangle
   =\frac{1}{n}\sum_{i\in A, j\in B}h_{|i-j|}, \notag \\
& c_3=\frac{1}{2}\langle\hat q_A\hat p_B+\hat p_B\hat q_A\rangle
   =\frac{1}{n}\sum_{i\in A, j\in B}k_{|i-j|}. \notag
\end{align}
As we do not observe the degrees of freedom of outside the regions A
and B, the evolution of this bipartite system is nonunitary. Thus, we take
into account the effect of decoherence through our definition of
bipartite system.  Using these components of the covariance matrix
$\bs{V}$, the symplectic eigenvalues are given by
\begin{align}
  &(\nu_{-})^2=a_1a_2-a_3^2+c_1c_2-c_3^2-|a_1c_2+a_2c_1-2a_3c_3|, \\
  &(\tilde\nu_{-})^2=a_1a_2-a_3^2-c_1c_2+c_3^2-|(a_1c_2-a_2c_1)^2
  +4(a_1c_3-a_3c_1)(a_2c_3-a_3c_2)|^{1/2},
\end{align}
and we can apply the separability criterion \eqref{eq:sepa-symp} to
judge whether the system is separable or entangled.

\subsection{Numerical result}
We calculated the logarithmic negativity of this system
numerically. The number of lattice sites is $N=100$ and the value of
the infrared cutoff parameter is chosen to be $\al=0.9999$.
\begin{figure}[H]
  \centering
  \includegraphics[width=0.45\linewidth,clip]{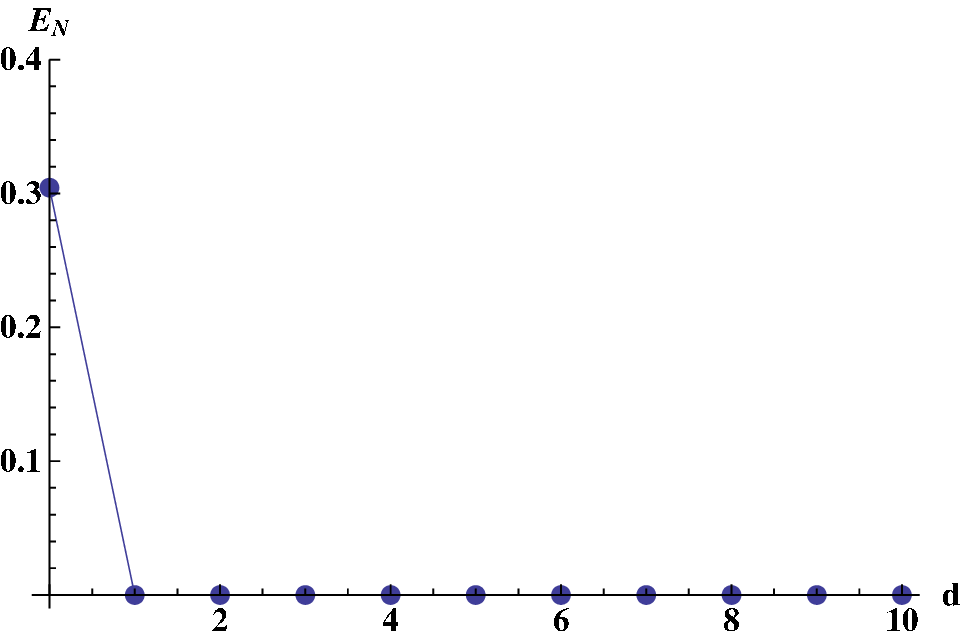}
  \includegraphics[width=0.45\linewidth,clip]{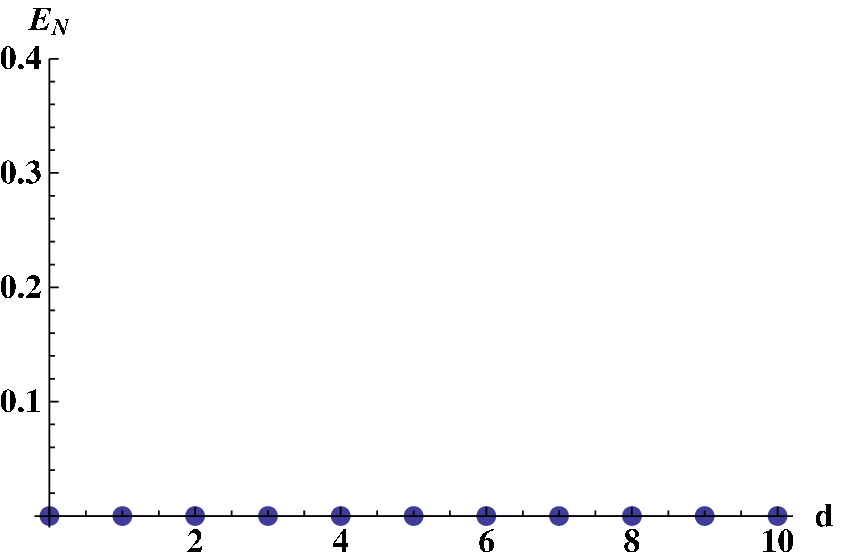}
  \caption{The dependence of the separation $d$ on $E_N$ for region size
    $n=4$. The left panel is $E_N$ at $\eta=-10$ and the right panel
    is $E_N$ at $\eta=-0.9$.}
  \label{fig:enta-distance}
\end{figure}
\noindent
Fig.~\ref{fig:enta-distance} shows the logarithmic negativity $E_N$ as
a function of separation $d$ between the regions A and B with the
region size $n=4$. Initially ($\eta=-10$, the left panel), $E_N\neq 0$
for $d=0$, and $E_N=0$ for $d\ge 1$. The regions A and B are entangled
for $d=0$ and separable for $d\ge 1$ . This implies that the system is
intrinsically entangled at this time because the choice of the
separation $d$ corresponds to the choice of measurement; how to
observe the system.  As the system evolves, the logarithmic negativity
becomes zero for any $d$ ($\eta=-0.9$, the right panel) and we can say
that the system becomes separable at this time. This behavior is 
not changed for the other value of the region size $n$. The spatial
structure of entanglement for this lattice model is simple and we only
pay attention to the behavior of entanglement for $d=0$.

\begin{figure}[H]
  \centering
  \includegraphics[width=0.5\linewidth,clip]{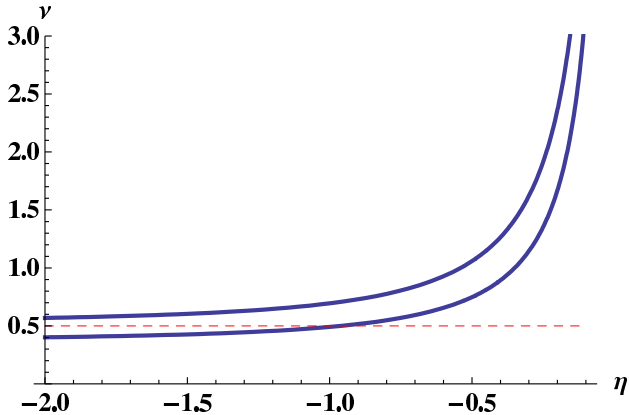}
  \caption{Evolution of symplectic eigenvalues for $d=0, n=4$. The
    upper line represents $\nu_{-}$ and the lower line represents
    $\tilde\nu_{-}$. The physical condition $\nu_{-}>1/2$ is always satisfied.}
  \label{fig:eigenvalue}
\end{figure}

Then, we consider the evolution of the entanglement.
Fig.~\ref{fig:eigenvalue} shows the evolution of symplectic
eigenvalues $\nu_{-}$ and $\tilde\nu_{-}$ for $d=0$ with $n=4$. During
the entire period of evolution, the value of $\nu_{-}$ is greater than
$1/2$ and the physical condition (\ref{eq:posi-symp}) is always
satisfied. On the other hand, the value of $\tilde\nu_{-}$ is smaller
than $1/2$ initially, then increases and exceeds the value
$1/2$. Thus, the initial entangled state changes into the separable
state.
\noindent
\begin{figure}[H]
  \centering
  \includegraphics[width=0.5\linewidth,clip]{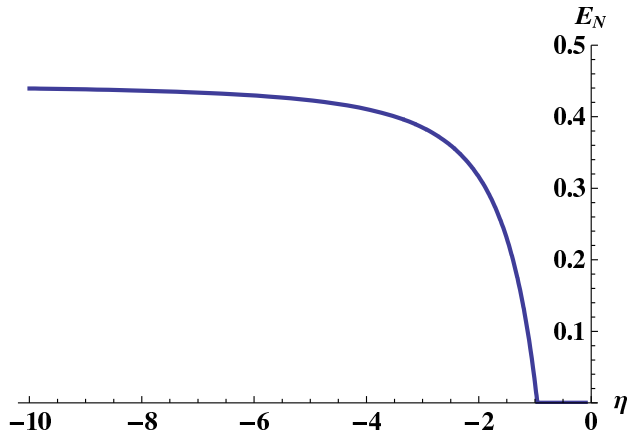}
  \caption{$E_N(d=0)$ as a function of the conformal time $\eta$
    for $n=4$ case. After $\eta=\eta_c\approx -1$, $E_N$ becomes zero
    and the system is separable.}
  \label{fig:enta-time}
\end{figure}
\noindent
We interpret these symplectic eigenvalues behaviors using the
logarithmic negativity.  Fig.~\ref{fig:enta-time} shows the
logarithmic negativity for $d=0$ as a function of conformal time. At
some critical time $\eta_c\approx -1$, the logarithmic negativity
$E_N$ becomes zero and the initially entangled state changes into a
separable state after $\eta_c$. As the quantum state, we assume the Bunchi-Davis
vacuum which imposes the Minkowski vacuum state in the short wave
length limit. Thus, the  entanglement of the scalar field before
$\eta_c$ implies the remnant of the entanglement of the Minkowski
vacuum.  After $\eta=\eta_c$, the regions A and B do not have
quantum correlation and we expect that the correlation between two
regions can be mimicked by an appropriate classical distribution
function. To understand what time scale determines the critical time
$\eta_c$, we observed how the critical time $\eta_c$ varies when we
change the region size $n$.
\begin{figure}[H]
  \centering
  \includegraphics[width=0.6\linewidth,clip]{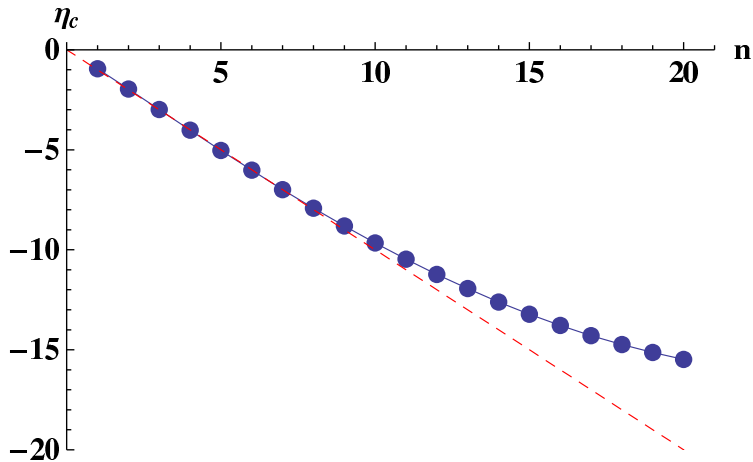}
  \caption{The relation between the critical time $\eta_c$ and the
    region size $n$. The deviation from a linear relation represents
    the effect of boundary condition. }
  \label{fig:critical-time}
\end{figure}
\noindent
Fig.~\ref{fig:critical-time} shows the relation between the critical
time $\eta_c$ and the region size $n$. For small value of $n$,
the relation coincides with the line $\eta_c=-n$ (dotted line). By restoring
the dimension of the variables, this relation corresponds to
\begin{equation}
  a(\eta_c)\times n\Delta x=\frac{1}{H}.
\end{equation}
Thus, the quantum entanglement between the regions A and B disappears
when the physical size of each region exceeds the Hubble horizon
length.  We must keep in mind that the physical size of each region
grows as $a(\eta)\times n\Delta x$ by the cosmic expansion. For larger
values of $n$, the relation deviates from the line $\eta_c=-n$ and we
suppose this is due to the effect of the periodic boundary condition
we imposed for the lattice model. The appearance of separability or
disentanglement at the Hubble horizon scale coincides with our naive
classical picture of the separate universe; the Hubble size
inflationary domains evolves independently and they can be treated
as independent Friedman-Robertson-Walker universes. As is shown here,
the quantum entanglement between the Hubble size regions is lost and
these regions do not have the quantum correlation. This is quantum
version of the separate universe picture. Hence, we can expect that the
quantum fluctuations in these regions behave classical.

\section{Entanglement and Quantum to Classical Transition}
To clarify the condition of quantum to classical transition of
fluctuations in the inflationary universe, we consider the relation
between the disentanglement and the classicalization of quantum
fluctuations. We can say that the quantum fluctuations become
classical if they are mimicked by appropriate classical stochastic
variables and their statistical nature cannot be distinguished from
quantum ones. More precisely, the bipartite quantum system under the
consideration is defined to be classical if there exists a positive
normalizable distribution function $\mathcal{P}$ on the phase space
and the following relation holds for any function $F$ of canonical
variables:
\begin{equation}
  \label{eq:classical-condition}
  \langle F(\hat q_A,\hat p_A,\hat q_B,\hat p_B)\rangle
  =\int dq_Adp_Adq_Bdp_B\mathcal{P}(q_A,p_A,q_B,p_B)F(q_A,p_A,q_B,p_B).
\end{equation}
The left-hand side of this relation is evaluated with respect to the
considering quantum state.  If this relation holds, any correlation
function of quantum variables in the bipartite system can be imitated
by the classical distribution function $\mathcal{P}$ on the phase
space and appropriate classical stochastic variables.

As is shown by Simon\cite{SimonR:PRL84:2000} and Duan \textit{et
  al.}\cite{DuanL:PRL84:2000}, for a bipartite two mode Gaussian
system, if the system is separable then the state of the system can be
written by the following form of $P$ representation:
\cite{GardinerCW:S:2004}
\begin{equation}
  \label{eq:P-repre}
  \hat\rho=\int d^2\al d^2\beta\, P(\al,\beta)|\al,\beta\rangle\langle\al,\beta|
\end{equation}
where $|\al,\beta\rangle=|\al\rangle|\beta\rangle$ is the product of
the coherent state of A and B, and $P(\al,\beta)$ is a positive
normalizable function called $P$ function. For a Gaussian state with
the covariance matrix $\bs{V}$, the $P$ function is also a Gaussian
function and given by
\begin{equation}
  \label{eq:P-function}
  P(\bs{\xi})=\frac{1}{4\pi^2}\sqrt{\mathrm{det}\bs{P}}\exp\left(-\frac{1}{2}
    \bs{\xi}^T\bs{P}\bs{\xi}\right), 
\qquad
  \bs{P}=\left(\bs{V}+\frac{1}{2}\bs{\Omega}\bs{S}^T\bs{S}\bs{\Omega}^T\right)^{-1}.
\end{equation}
where $\bs{S}\in\mathrm{Sp}(2,R)\otimes\mathrm{Sp}(2,R)$ is the
symplectic transformation that transforms the covariance matrix
$\bs{V}$ to the following standard form\cite{DuanL:PRL84:2000}:
\begin{equation}
  \label{eq:S-transf}
  \bs{V}_{II}=\bs{S}\bs{V}\bs{S}^T=
  \begin{pmatrix}
    a r & & c r & \\ & a/r && c'/r\\ c r&& ar & \\ & c'/r && a/r
  \end{pmatrix}
, \quad r=\sqrt{\frac{a-|c'|}{a-|c|}}.
\end{equation}
From the definition of $P$ function (\ref{eq:P-function}), we have
\begin{equation}
  \bs{P}=\bs{S}^T\left(\bs{V}_{II}-\frac{\bs{I}}{2}\right)^{-1}\bs{S}
\end{equation}
and the existence of a positive normalizable $P$ function is guaranteed by the
condition $(\bs{V}_{II}-\bs{I}/2)\ge 0$. In terms of the components
of $\bs{V}_{II}$, this condition is represented as 
\begin{equation}
  (a-|c|)(a-|c'|)\ge \frac{1}{4}.
\end{equation}
On the other hand, the symplectic eigenvalues of $\bs{V}$, which is
invariant under symplectic transformations, are
\begin{align}
  &\nu^2=
  \begin{cases}
    (a-|c|)(a-|c'|), (a+|c|)(a+|c'|),\quad & cc'\ge 0\\
    (a-|c|)(a+|c'|), (a+|c|)(a-|c'|),\quad & cc'<0
  \end{cases} \\
  &\tilde\nu^2=
  \begin{cases}
    (a-|c|)(a+|c'|), (a+|c|)(a-|c'|),\quad & cc'\ge 0 \\
    (a-|c|)(a-|c'|), (a+|c|)(a+|c'|),\quad & cc'<0.
  \end{cases}
\end{align}
Therefore, the condition of existence of positive normalizable
$P$ function is equivalent to the separability condition
(\ref{eq:sepa-symp}) provided that the physical condition
(\ref{eq:posi-symp}) is satisfied.  From the definition of
$P$ representation (\ref{eq:P-repre}), if the system is separable, it
is possible to calculate the quantum expectation value of the normally
ordered product of any operators using the $P$ function as a distribution function
\begin{equation}
  \label{eq:normal-ordering}
  \langle : F(\hat q_A,\hat p_A,\hat q_B,\hat p_B):\rangle=\int
  dq_Adp_Adq_Bdp_B P(q_A,p_A,q_B,p_B)F(q_A,p_A,q_B,p_B).
\end{equation}
However, the existence of $P$ function is not sufficient for the
establishment of classicality of the system; it only guarantees the
existence of the distribution function for the normally ordered
quantities.

To derive the condition of classicality of the quantum field, we
introduce a Wigner distribution function on the phase space and its
form for a Gaussian state is given by
\begin{equation}
  \label{eq:wigner}
  W(\bs{\xi})=\frac{1}{4\pi^2\sqrt{\mathrm{det}\bs{V}}}
  \exp\left(-\frac{1}{2}\bs{\xi}^T\bs{V}^{-1}\bs{\xi}\right).
\end{equation}
The normalizable Winger function exists for $\bs{V}\ge 0$ and this
condition is weaker than the physical condition of the state
\eqref{eq:positivity}. Hence, there exists a normalizable Wigner
function, which does not represent the physical state. The Wigner
function (\ref{eq:wigner}) for $\bs{V}\ge 0$ is positive normalizable
and can be interpreted as a distribution function giving
the expectation value for the symmetrically ordered product of
operators
\begin{equation}
  \label{eq:symmetrized}
  \langle\left\{F(\hat q_A,\hat p_A,\hat q_B,\hat p_B)\right\}_{\text{sym}}\rangle=\int
  dq_Adp_Adq_Bdp_B W(q_A,p_A,q_B,p_B)F(q_A,p_A,q_B,p_B).
\end{equation}
If the difference between the $P$ function and the Wigner function is
negligible, these distribution functions return the same answer for
the expectation value of any operator $\hat F$ and we have the relation
\begin{equation}
  \label{eq:classicality}
  \langle:\hat F:\rangle\approx  \langle\{\hat
  F\}_{\text{sym}}\rangle\approx  \langle\hat F\rangle.
\end{equation}
This means that noncommutativity between operators is negligible and
the $P$ function or the Wigner function plays a role of the classical
distribution function, which reproduces the quantum expectation value
for any operators. Hence, the condition of classicality
\eqref{eq:classical-condition} is established.

We look for the condition for the establishment of the relation
(\ref{eq:classicality}). For this purpose, it is sufficient to
consider the condition for the standard form of the covariance matrix
$\bs{V}_{II}$ because this form of the covariance matrix is related to
the original covariance matrix $\bs{V}$ via a symplectic
transformation. In terms of the covariance matrix, the condition for
the classicality $P\approx W$ is given by
\begin{equation}
  \label{eq:classicality2}
  \bs{V}_{II}^{-1}\approx (\bs{V}_{II}-\bs{I}/2)^{-1}.
\end{equation}
We write down the components of $\bs{V}_{II}^{-1}$ and
$(\bs{V}_{II}-\bs{I}/2)^{-1}$ explicitly
\begin{align}
  &\bs{V}_{II}^{-1}=\begin{pmatrix}
    \dfrac{a/r}{a^2-c^2} & & -\dfrac{c/r}{a^2-c^2} &
    \\
    & \dfrac{ar}{a^2-c'^2} & & -\dfrac{c'r}{a^2-c'{}^2} \\
    -\dfrac{c/r}{a^2-c^2}&&\dfrac{a/r}{a^2-c^2} & \\
    &-\dfrac{c'r}{a^2-c'^2}&&\dfrac{ar}{a^2-c'^2}
    \end{pmatrix}, \notag \\
    &
    (\bs{V}_{II}-\bs{I}/2)^{-1} \notag\\
&=
    \begin{pmatrix}
      \dfrac{(a-1/2r)/r}{(a-1/2r)^2-c^2} & &
      -\dfrac{c/r}{(a-1/2r)^2-c^2} & \\
     & \dfrac{(a-r/2)r}{(a-r/2)^2-c'^2} & & -\dfrac{c'r}{(a-r/2)^2-c'^2}\\
     -\dfrac{c/r}{(a-1/2r)^2-c^2}&&\dfrac{(a-1/2r)/r}{(a-1/2r)^2-c^2}& \\
     &-\dfrac{c'r}{(a-r/2)^2-c'^2}&&\dfrac{(a-r/2)r}{(a-r/2)^2-c'^2}
    \end{pmatrix}. \notag
\end{align}
Thus, if the condition
\begin{equation}
  \label{eq:cond-classical}
  f_1\equiv a^2r^2=a^2\frac{a-|c'|}{a-|c|}\gg\frac{1}{4},\quad 
  f_2\equiv \frac{a^2}{r^2}=a^2\frac{a-|c|}{a-|c'|}\gg\frac{1}{4}
\end{equation}
is satisfied, $\bs{V}_{II}^{-1}\approx (\bs{V}_{II}-\bs{I}/2)^{-1}$
and the $P$ function equals  the Wigner function. We rewrite the
condition \eqref{eq:cond-classical} in terms of the symplectic
eigenvalues. For $cc'>0$, $\nu_{-}^2=(a-|c|)(a-|c'|)$ and
\begin{equation}
  f_1=\frac{a^2\nu_{-}^2}{(a-|c|)^2}>\nu_{-}^2,\quad
  f_2=\frac{a^2\nu_{-}^2}{(a-|c'|)^2}>\nu_{-}^2,\quad
\end{equation}
and the condition \eqref{eq:cond-classical} is satisfied if
$\nu_{-}\gg 1/2$. For $cc'<0$, 
\begin{equation}
  f_1=\frac{a^2\tilde\nu_{-}^2}{(a-|c|)^2}>\tilde\nu_{-}^2,\quad
  f_2=\frac{a^2\tilde\nu_{-}^2}{(a-|c'|)^2}>\tilde\nu_{-}^2,\quad
\end{equation}
and the condition \eqref{eq:cond-classical} is satisfied if
$\tilde\nu_{-}\gg 1/2$. Combining these two cases, we have the
following result:
\begin{equation}
  \nu_{-}\gg \frac{1}{2},\quad \tilde\nu_{-}\gg\frac{1}{2}\quad
  \Longrightarrow\quad \bs{V}_{II}^{-1}\approx
  (\bs{V}_{II}-\bs{I}/2)^{-1}\quad \Longleftrightarrow\quad P\approx W.
\end{equation}
Therefore, $\nu_{-},\tilde\nu_{-}\gg 1/2$ is the sufficient condition for
the system can be treated as classical.

We can check whether this condition is satisfied in our lattice
model. As is show in Fig.~\ref{fig:eigenvalue}, before the critical
time $\eta_c\approx -1$, $\nu_{-}>1/2$ and $\tilde\nu_{-}<1/2$ and the
system is entangled. After $\eta_c$, the value of $\tilde\nu_{-}$
becomes greater than 1/2 and increases in time. In our lattice model, the
relation $\tilde\nu_{-}<\nu_{-}$ always holds. For $-1\lesssim
\eta< 0$, the behavior of $\tilde\nu_{-}$ is approximately
given by
\begin{equation}
  (\tilde\nu_{-})^2-\frac{1}{4}\approx \frac{0.10605}{\eta^2}-0.115144+O(\eta^2)
\end{equation}
and after $\eta=\eta_c$, the condition $\tilde\nu_{-}\gg 1/2$ is
rapidly realized. Hence, the difference between the $P$ function and
the Wigner function becomes negligible in one Hubble time after the
system becomes separable at the horizon crossing. As a subset of the
separability condition for $\bs{V}_{II}$, the inequality
\begin{equation}
  a^2\ge\frac{1}{4}
\end{equation}
holds for the physical state and this corresponds to a standard
uncertainty relation. The condition of classicality
$\nu_{-},\tilde\nu_{-}\gg 1/2$ leads to $a^2\gg 1/4$ and this also
implies the noncommutativity between canonical variables can be
neglected when we evaluating the expectation values of operators. This
is consistent with the result obtained in the paper
\cite{GuthAH:PRD32:1985,AlbrechtA:PRD50:1994,PolarskiD:CQG13:1996,LesgourguesJ:NPB497:1997,KieferC:CQG15:1998};
for superhorizon scale quantum fluctuations, the noncommutativity
between canonical variables becomes negligible because the growing
mode solution is dominant. In other words, we can neglect $\hbar$ in
the uncertainty relation. We derived the equivalent condition for the
classicality from the condition of the existence of the classical
distribution function and the symplectic eigenvalues.

\section{Summary and Conclusion}

We investigated the appearance of the classical distribution function
for the quantum fluctuation in the inflationary model using the
lattice model of the scalar field. By following the evolution of
entanglement between two spatially separated regions, we found the
classicality of the quantum field appears as follows: Initially, when
the size of the considering region is smaller than the Hubble horizon,
the quantum field is in the entangled state. As the Universe expands,
the quantum state becomes separable when the size of the region equals
 the size of the Hubble horizon. At this stage, the quantum correlation
between neighboring regions is lost.  Then, within about one Hubble
time after the horizon crossing, noncommutativity of operators
becomes negligible and the system can be treated as classical.
Any quantum expectation values can be evaluated using the $P$ function or
the Wigner function. In other words, there appears classical
stochastic nature for variables, which mimics the original quantum
dynamics.  As we have shown, disentanglement is not a sufficient
condition for the establishment of classicality of fluctuations. This
condition only guarantees the loss of EPR-type nonlocal correlations,
which are peculiar to quantum mechanics.

In our analysis, we defined a bipartite system as the subsystem of the
entire universe and we discard the unobserved dynamical degrees of
freedom outside of the observed region. Thus, our bipartite system
evolves in a nonunitary way and this definition of our system effectively
takes into account the decoherence mechanism of the considering
region.

We comment on the relation of our analysis to the stochastic approach
of inflation\cite{StarobinskiA:1986} which treats the quantum dynamics
of inflaton fields as the classical stochastic process.  By coarse
graining the scalar field on the large scale $(\ep H)^{-1}, \ep\ll 1$,
it can be shown that the coarse grained field obeys the Langevin
equation and the dynamics of the quantum inflaton field is replaced by
the classical stochastic process. In the stochastic approach, the
classical nature of the inflaton field is guaranteed by the
appropriate small value of the coarse graining parameter
$\epsilon$. However, in this approach, the connection between the
probability distribution and the state of the inflaton field is not
clear.  The stochastic approach assumes the existence of the classical
probability distribution from the first. However, as we have shown in
this paper, it is possible to define a probability distribution
function only when the system becomes separable. We expect that the
condition $\epsilon\ll 1$ corresponds to the condition $\tilde\nu\gg
1/2$ which is stronger than the separability. Anyways, we must
reconsider the meaning of the probability in the stochastic approach
from the view point of entanglement. We will report on this topic in
a separate publication.

In this paper, we assumed the Bunch-Davis vacuum state. Previous
analysis by J. Lesgourgues \textit{et
  al.}\cite{LesgourguesJ:NPB497:1997} considered the nonvacuum initial
states that are non-Gaussian and concluded that the non-Gaussian
nature of the state does not affect the transition to the classical
behavior. However, from the viewpoint of the entanglement, the
condition of the separability for non-Gaussian states is unknown and
the determination of a classicality condition for such states is an
unsolved problem. Further, we considered the quantum to classical
transition based on the bipartite entanglement only. This is because
the criterion on the separability for the general $N$-partite system
is unknown\cite{BraunsteinSL:RMP77:2005}. However it is necessary to
look for the classicality condition for non-Gaussian states and the 
$N$-partite system to fully understand the mechanism of classical to
quantum transition of primordial fluctuation. This is a future
problem to be tackled.

\begin{acknowledgments}
The author would like to thank Yuji Ohsumi for valuable discussions on this subject.
This work was supported in part by a JSPS Grant-In-Aid for Scientific
Research [C] (19540279).
\end{acknowledgments}


\end{document}